\documentclass[11pt,preprint]{aastex}
\usepackage{graphics}
\usepackage{amssymb}
\usepackage{amsmath}
\usepackage[dvips]{color}
\newcommand{\kms}{{km~s$^{-1}$}}
\newcommand{\fex}{[Fe {\sc X}]$\lambda$6376}

\newcommand{\fexiv}{[Fe {\sc XIV}]$\lambda$5304}
\newcommand{\fevii}{[Fe {\sc VII}]}

\newcommand{\oiii}{[O {\sc III}]$\lambda$5007}
\newcommand{\oiiis}{[O {\sc III}]$\lambda$4959}
\newcommand{\neiii}{[Ne {\sc III}]$\lambda$3896}
\newcommand{\hb}{H$\beta$}
\newcommand{\heii}{He {\sc II}$\lambda$4686}

\newcommand{\ha}{H$\alpha$}

\shortauthors{Yang et al.}
\shorttitle{Long Term Evolution of TDE Candidates}

\begin{document}
\title{Long Term Spectral Evolution of Tidal Disruption 
Candidates Selected by Strong Coronal Lines}
\author{Chen-Wei Yang\altaffilmark{1}, Ting-Gui Wang\altaffilmark{1}, Gary Ferland\altaffilmark{2}, 
Weimin Yuan\altaffilmark{3}, Hong-Yan Zhou\altaffilmark{1,4}, Peng Jiang\altaffilmark{1}}
\altaffiltext{1}{Key Laboratory for Research in Galaxies and Cosmology, The University of Sciences and 
Technology of China, Chinese Academy of Sciences, Hefei, Anhui 230026, China (twang@ustc.edu.cn)}
\altaffiltext{2}{Department of Physics, University of Kentucky, Lexington, KY 40506, USA}
\altaffiltext{3}{National Astronomical Observatory, Chinese Academy of Sciences,20A Datun Road, Beijing, China}
\altaffiltext{4}{Polar Research Institute of China, 451 Jinqiao Road, Pudong, 
Shanghai 200136, China}

\begin{abstract}
We present results of follow-up optical spectroscopic observations
of seven rare, extreme coronal line emitting galaxies reported by Wang et al. (2012) 
with Multi-Mirror Telescope (MMT). 
Large variations in coronal lines are found in four objects, making them strong candidates of tidal disruption events (TDE). 
For the four TDE candidates, all the coronal lines with ionization status higher than \fevii\ 
disappear within 5-9 years.
The \fevii\ faded by a factor of about five in one object (J0952+2143) within 4 years, 
whereas emerged in other two without them previously. 
A strong increment in the [O {\sc III}] flux is observed, 
shifting the line ratios towards the loci of active galactic nucleus on the BPT diagrams. 
Surprisingly, we detect a non-canonical \oiii/\oiiis$\simeq$2 in two objects,
indicating a large column density of O$^{2+}$ and thus probably optical thick gas.
This also requires a very large ionization parameter and relatively 
soft ionizing spectral energy distribution (e.g. blackbody with $T < 5\times 10^4$~K).
Our observations can be explained as echoing of a strong ultraviolet to soft X-ray flare caused by tidal disruption events, on molecular clouds in the inner parsecs of the galactic nuclei. Re-analyzing the SDSS spectra reveals double-peaked or strongly blue-shouldered broad lines in three of the objects, which disappeared in the MMT spectra in two objects, and faded by a factor of ten in 8 years in the remaining object with a decrease in both the line width and centroid offset. We interpret these broad lines as arising from decelerating biconical outflows. 
Our results demonstrate that the signatures of echoing can persist for as long as ten years, 
and can be used to probe the gas environment in the quiescent galactic nuclei.
\end{abstract}


\section{Introduction}

Stellar and gaseous kinematics suggests that most galaxies harbor central 
supermassive black holes (SMBH) and that the SMBH masses are well correlated with the 
stellar masses or velocity dispersions of the galactic bulges (e.g. Kormendy \& 
Richstone 1995; Magorrian et al. 1998; Ho 1998; Merritt \& Ferrarese 2000; 
Gebhardt et al. 2000). When a star in the galactic nucleus accidentally passes 
within the tidal disruption radius of the massive black hole ($r_p < R_T 
\simeq R_*(M_{\rm BH}/M_*)^{1/3}$, Hills 1975), it is torn apart by the tidal 
force of the black hole. About half of the stellar debris is ejected and the 
rest falls back towards the black hole, causing a bright flare lasting for a 
few months to years (Rees 1988; Ayal et al. 2000). For a black hole of mass 
between $10^7\ M_{\sun}$ and $10^8\ M_{\sun}$ and a solar-type star, the peak 
luminosity will be sub-Eddington ($L_{Edd}=1.3\times 10^{38} (M_{\rm BH}/M_\sun) 
{\rm  erg\ s}^{-1}$). For less massive black hole ($M_{\rm BH} < 10^7\ M_{\sun}$), 
the rate of fall back initially exceeds the Eddington accretion rate, likely 
launching strong outflows. The spectra of the super-Eddington accretion tori and outflows are still uncertain, but when the fallback rate is below the Eddington rate, the gas is accreted onto the black hole via a thin disk and its radiation peaks in the UV to soft X-ray bands.

Besides UV and soft X-ray continuum flares, variable broad and narrow emission lines 
may be also seen in the spectra. Part of the UV and X-ray light may be reprocessed by 
the outflows or unbounded debris, giving rise to the broad emission lines (Bogdanovi\'c 
et al. 2004; Strubbe \& Quataert 2009; Wang et al. 2011, hereafter W11; Gezari et al. 
2012). The same UV and X-ray sources may also ionize the cold circum-nuclear interstellar 
medium (ISM), producing high ionization narrow lines (Ulmer 1999; Komossa et al. 2008; 
Wang et al. 2011, 2012). Broad lines are expected to follow the continuum 
variations closely because they are produced in a region very close to the continuum source.
Narrow lines should also be variable, but on a longer time scale due to the response 
of ISM material further away to the continuum flare (light echoing). In particular, the 
 high-ionization narrow lines which vary on time scales of years are unique features for the 
light echo of tidal disruption event (TDE), that distinguishes it from active galactic 
nuclei (AGNs), which possess narrow line regions (NLRs) of sizes from $10^2$ to $10^3$ 
pc. Because gas at large scale will have little response to the continuum flare on time 
scales of months to years, we may expect to observe only the spectrum from the very inner 
region of the NLR, which is likely dominated by high ionization lines, while low ionization 
lines from normal NLR is weak.

Wang et al. (2012, hereafter W12) carried out a systematic search for extreme coronal 
line emitters (ECLEs) in the spectroscopic samples of galaxies and low redshift quasars 
from the Sloan Digital Sky Survey (SDSS) Data Release 7 (DR7), following the initial 
results of detecting variable narrow lines in two ECLEs (Komossa et al. 2008, 2009; 
W11). They identified a sample of seven galaxies that show extremely strong coronal 
lines from [Fe {\sc X}] up to [Fe {\sc XIV}]. Traditional narrow-line diagnostics 
suggested they are non-active, and half of the sample show broad emission lines of 
complex profiles. Also the sample is split into low and high ionization subclasses 
according to the detection or non-detection of \fevii. W12 showed that these lines are formed in 
a photo-ionized gas, and argued that ECLEs are most likely the light echoes of 
a tidal disruption flare based on  energetics considerations. Most these galaxies are  
intermediate luminosity disk galaxies with small bulges. If all these sources  
turn out to  be TDE, then the rate of such TDE is expected around a few times $10^{-5}$ per galaxy 
per year for galaxies in the absolute magnitudes around $-21.3 <M_i< -18.9$. 

However, there are some further questions left from previous studies. First, optical follow-ups 
were carried out for only two of the seven objects. It remains to be verified whether the rest of 
the sample also show the same variation trends. Second, the long term evolution of the emission 
lines is not known. On what time scale do coronal lines disappear? Are there any other special 
features in the later stage of evolution that can be used to identify such events on long time 
scales? With these questions, we initiated spectroscopic follow-up observations with the 
Multi-Mirror telescope (MMT) to characterize the spectral signatures at  
late phases of evolution. In this paper we will present an analysis of the follow-up MMT 
spectroscopic observations. The paper is organized as follows. We present the observations and data 
reduction in \S \ref{sec:observations} and results on the spectral evolution in \S \ref{sec:results}.
We discuss the implication of these results in \S \ref{sec:discussion}. Throughout the paper, we will 
adopt a $\Lambda$-CDM cosmology with $H_0 = 71 {\rm km\ s}^{-1}\ {\rm Mpc}^{-1}$, $\Omega_M = 0.28$ and
$\Omega_\Lambda= 0.72$.

\section{Observations and Data Reduction\label{sec:observations}}

We observed all  seven targets with the Blue Channel Spectrographs mounted on the 
Multi-Mirror Telescope (MMT) on December 26, 2011. The log of the observations 
is given in Table 1. During the observation, the typical seeing was 0$\farcs$8 and 
we adopt a long slit with 1$\arcsec$ width. The slit is centered on the galactic 
nuclei and orientated at the parallactic angle. We took exposures of all  seven 
targets for 900s with 500 gpm grating, centering at 6000\AA. This setting results 
in a spectral resolution $R=\lambda/\Delta\lambda=1430$ and a wavelength coverage 
from 4430\AA\ to 7560\AA. Unfortunately, the slit was not properly placed on the 
center of the galaxy for SDSS J1241+4426, thus we will ignore the red part of 
spectrum for this object. We also obtained the blue portion of spectra for five 
objects, including SDSS J0748+4712, SDSS J0938+1353, SDSS J0952+2143, 
SDSS J1055+5637, and SDSS J1241+4426 using 800 gpm grating, centered at 
4100{\AA}, and an exposure time of 900s for each. The latter gives a spectral resolution 
of 1730 and a wavelength coverage from 3100\AA\ to 5100{\AA}. Three and one KNPO 
standard stars were observed using the same settings as 500 gmp and 800 gmp during 
the observation for flux calibration, and 7 and 5 He-Ne-Ar lamp spectra with g500 and 
g800 grism sets were taken for the wavelength calibration. The raw two-dimension data 
reduction and spectral extraction were accomplished using standard routines in IRAF. 
To extract the nuclear spectra, we used the APALL task\footnote{APALL is a multi-step 
task which defines and extracts the data from 2D CCD image in IRAF.} and choosed an 
aperture of 3$\arcsec$. We also extract the spectra with an aperture of 1$\arcsec$, all 
high ionization lines including coronal lines, [O {\sc III}] and \heii\ are almost 
the same, so must come from the center 1$\arcsec$ region. We carried out the 
wavelength and flux calibration using the spectra of a He-Ne-Ar lamp and those of the 
standard stars, respectively. When applying different 
standard stars for flux calibration taken with 500 gpm, the spectra of the same object 
were consistent with each other within 5\%. Thus, we take the medians as our final results. 
After the flux calibration, the blue and red spectra of the same object taken with 
the two gratings are consistent with each other within 8\% in the overlapping region, 
indicating small uncertainty in the flux calibration. Then we rescale the blue part 
spectrum to match the red one, and then combine them. We estimate the spectral 
resolution by measuring the width of emission lines in the spectrum of He-Ne-Ar 
lamp. For 500 gpm grating and 800 gpm grating, their typical values are 90 \kms\ and 69 \kms.

\section{Results\label{sec:results}}

\subsection{Comparison between SDSS and MMT spectra}

The MMT spectra are shown in Figure 1. For comparison, we also overlay the SDSS 
spectra taken 5-9 years ago. Note that the SDSS spectra were taken using a larger 
aperture, a circle of 3$\farcs$0 in diameter, than ours, in 1$\farcs$0 $\times$ 3$\farcs$0 
region, so they contain more light from the host galaxy than our MMT spectra. As such the 
low-ionization narrow lines from the extended star-formation regions may be significantly 
smaller in the MMT spectra, along with the starlight; while we do not expect that a 
significant coronal lines come from the extended region. Therefore, coronal lines 
will not be significantly affected by the aperture effect. 

An initial check of the Figure 1 suggests two different variability trends for 
coronal lines. In four of seven objects, the high-ionization coronal lines \fex-\fexiv\  
completely disappeared in the MMT spectra, while for other three objects 
J0938+1353, J1055+5637 and J1241+4426, the MMT spectra are similar to the 
SDSS ones with strong high-ionization coronal lines. For J1241+4426, 
although we failed to obtain the red portion spectrum of the galactic center, 
we find that [Fe {\sc VII}]$\lambda$3759 and \neiii\ lines are prominent and 
do not show significant variation in the blue portion of the spectrum. Therefore, 
the latter three objects seem to be AGNs. The broad Balmer lines, Fe {\sc II} and the 
non-stellar continuum remain very prominent in J1055+5637, thus it is a type 
1 Seyfert galaxy. Note that J1055+5637 is the only object in W11 with narrow-
line ratios located in the AGN regime in all three BPT diagrams. J0938+1353 shows 
fairly broad coronal and high ionization lines, superposing on strong, much 
narrower, low ionization lines, which come clearly from star-formation regions.
We suspect that coronal lines and high ionization lines are from an obscured AGN. 
Thus it is likely a composite of a Seyfert 2 nucleus with star-formation regions but its true nature remains mysterious as coronal lines are not usually seen in composite type galaxies. Alternatively, the 
coronal lines may be related to tidal disruption of a giant star by a massive black hole, which produces a much longer lasting flare than of tidal disruption of a main sequence star. The latter scenario can be tested by future spectroscopic follow-up observations. We will focus only on the four variable 
objects in the late analysis. 

\subsection{Continuum Modeling and Variability}

In this subsection we will model the continuum and examine the contribution 
of potential non-stellar continua in the SDSS spectrum. W12 found that four of 
the seven objects show flux variations between SDSS photometric and spectral 
observations, including the AGN J1055+5637. For two objects without \fevii\ 
lines, they were brighter during the spectroscopic observations than during the 
photometric observation, indicating that there is a non-stellar continuum in 
J0748+4712 and J1350+2916 spectra, while it is difficult to assess the case in J0952+2143. 
If TDE is responsible for the continuum variability, any non-stellar continuum 
emission would be very weak after 5-9 years even if it existed during the SDSS 
spectroscopic observations. Indeed, in all MMT spectra, the continua are dominated 
by starlight. Therefore, the continua in the MMT spectra can be used as first-
order templates for the stellar light in the SDSS by ignoring the radial gradients of 
stellar populations. 

In order to get the starlight spectrum  beneath the emission lines and also beyond 
the MMT spectral coverage, we fit the MMT spectra with a combination of 6 independent 
components (ICs) templates derived from the stellar populations (BC03) with Ensemble 
Learning Independent Component Analysis. Previous tests show that such extrapolation 
is reliable (see Lu et al. 2006 for detail). In practice, the continuum fit to a narrow 
wavelength coverage (4430\AA\ to 7560\AA ) of the MMT spectrum of J1350+2916 can match 
its SDSS spectrum (wavelength coverage from 3700\AA\ to 8500\AA) well. During the fitting 
we mask all prominent emission lines, and convolve the templates with a Gaussian kernel 
and shift in redshift to match the width and the centroid of the absorption lines. The 
fits with 6 ICs are shown also in Figure 1.
 
Keeping in mind that different amounts of the starlight lie within the apertures, we rescale 
the starlight models of J0748+4712, J0952+2143 and J1350+2916 derived from the MMT 
spectra to match their SDSS ones by minimizing the residuals in the stellar absorption 
lines. This yields the fit in Figure 2. Overall, the stellar absorption lines in the SDSS 
spectra are fitted very well, suggesting that this approach is feasible. A non-stellar 
continuum is required to be present in only SDSS J0748+4712. This provides an independent 
confirmation to the continuum variability analysis, further it also gives an estimate of the 
non-stellar continuum spectrum. Note this is much less model dependent than the spectral 
decomposition method using a combination of starlight plus a power-law or black-body fit 
in W11. We will describe the individual object below.  

For J0748+4712, its SDSS spectrum shows strong non-stellar emission 
and three broad bumps around 4050, 4600 and 6560\AA\ (see W11). The scaling 
factor of the star-light in the SDSS spectrum is 1.35. After subtracting the 
stellar light spectrum, the differential spectrum is a smooth continuum plus 
several broad bumps and narrow emission lines (Figure 2). One  noticeable 
difference from W11 is that there is no broad feature around 4050\AA, suggesting 
it was due to imperfect subtraction of stellar light in W11. The continuum is 
blue, but there is also a hint of curvature in the non-stellar spectrum. The 
latter is consistent with the variability analysis in W12 that the amplitudes 
appears larger in $g$ and $i$-bands than in $r$ band. But we cannot rule out the 
possibility that there is an additional old stellar population in the SDSS 
spectrum. If the curvature is confirmed, the non-stellar continuum may consist 
of two components, e.g., a disk component and a reprocessed outflow component. 

It was reported that J0952+2143 also has a non-stellar continuum in 
the SDSS spectrum (Komossa et al 2009). These authors fitted the SDSS spectrum 
with a single simple stellar population plus a black-body emission. Our analysis 
does not confirm this. The SDSS continuum can be fitted very well with a scaled 
up MMT spectrum, therefore, a non-stellar component is not needed. For J1350+2916, 
W12 found that there is 3$\sigma$ evidence for brightening in the $g$ band between 
SDSS photometric and spectroscopic observations. Our analysis does not require 
a non-stellar component in this case. However, for this object, the MMT spectrum 
does not cover the 4000\AA\ break or blueward of it, which is crucial to the 
detection of non-stellar component, because a non-stellar continuum is most 
prominent in short band and also the 4000\AA\ break provides strong constraints on 
the scale of the stellar component.

\subsection{Broad Emission Line Variability}

Three of the four objects displayed prominent broad \ha, \hb\ or \heii\ 
lines in their SDSS spectra. Broad Balmer emission lines in J0952+2143 
and J1350+2916 displayed double horns with separations between 2000 -- 
3000 \kms (see Komossa et al. 2008; W12). The FWHMs of these lines are 2100 
and 2600 \kms, respectively. J0748+4712 showed two strong and 
broad (several hundred \AA) bumps peaked around 4600 and 6560\AA. The broad 
bump at 4600\AA\ was interpreted as blue-shifted He {\sc II} by W11. With the 
more reliable starlight subtraction here, this line also seems  double peaked 
with the red peak close to the rest frame of He {\sc II} while the blue peak 
at $-8,000$ \kms, although the signal to noise ratio is still a bit low. It is
hard to attribute the blue peak to the contamination 
of another broad line as there is no strong lines expected in this wavelength 
range. Although the double peaks are not evident in the broad {\ha}
due to its weakness, the data are consistent with this. Thus, broad lines in 
all three TDE candidates show double peaked profiles. We have fitted the narrow \ha\ 
and [N {\sc II}] double lines with three narrow Gaussians and use a broad Gaussian 
to estimate the broad component of \ha\ line. \ha\ is detected with a flux of 
2.1$\times 10^{-15}$erg~cm$^{-2}$~s$^{-1}$. This gives a He {\sc II}/{\ha} ratio 
of 3.3. 

These broad bumps in J0748+4712 and double horns in J0952+2143 and J1350+2916 
disappeared in the MMT spectra (Figure 3). Only very weak broad H$\alpha$ and 
H$\beta$ lines can be spotted in the MMT spectrum of J0952+2143 (middle panel 
of Figure 3 and figure 4). Any broad lines, if present, must be below the 
detection limits in the other two objects. 

To measure the broad emission lines in J0952+2142, we fit the continuum 
subtracted spectra using Gaussians (Figure 4). All narrow lines show symmetric 
profiles.  The \ha\ and \hb\ lines display an additional redshifted broad component. 
Each symmetric narrow line is fitted with a Gaussian, while one more 
Gaussian component is added for the redshifted broad component of each Balmer 
line. \ha\ and \hb\ are fitted, with the widths and centroids of each component 
locked. The line flux are listed in table 2. The broad component is redshifted by about 120 \kms\ 
relative to the systematic velocity and has an FWHM of 620$\pm$45 \kms. It could 
be the relic of the fading broad Balmer lines seen in the SDSS and NTT spectra 
that was described in Komossa et al. (2008 \& 2009). In those two spectra, 
which were taken on December 30, 2005 and Feb 6, 2008, the broad Balmer 
components had redshifts of 560 \kms\ and 270 \kms, and FWHMs of 2100\kms\ 
and 1500\kms. The line flux in the MMT spectrum is only 9\% of that in SDSS 
spectrum and 25\% in NTT spectrum.   

\subsection{Coronal Line Variability}

Figure 5 compares the coronal line spectra, after subtracting the continuum of MMT 
spectra from those of the SDSS spectra over the wavelength range from 5200\AA -6450\AA, 
covering coronal lines {\fevii}, {\fex} and {\fexiv}. The one Gaussian fit result are also listed in table 2. A close look at the figure reveals some 
interesting variation patterns. Two (J1342+0530 and J1350+2916) of three objects 
without \fevii\ emission lines in the SDSS spectra now show these lines in the MMT spectra, 
while in J0952+2142, \fevii\ lines prominent in the SDSS and NTT spectra become very weak 
now. But individual objects also show some different properties, that will 
be further described below.

J0748+4712 -- its SDSS spectrum shows strong high-ionization coronal lines 
([Fe {\sc X}], [Fe {\sc XI}] and [Fe {\sc XIV}]) but no [Fe {\sc VII}] lines. In the 
follow-up spectroscopic observation in 4-5 years after the SDSS observation, all 
the coronal lines disappeared although the S/N ratios of the spectra were very low 
(W11). The high S/N ratio MMT spectrum confirmed this, and there were no \fevii\ 
or higher ionization lines. This marks this object with the shortest duration of 
coronal lines.

J0952+2143 -- its SDSS spectrum shows both high-ionization coronal lines and 
relative low-ionization coronal lines ([Fe {\sc VII}]). In the follow-up 
observations carried out with the Xinglong 2.16m telescope and the ESO NTT telescope 
2-3 years later, all the coronal lines with ionization higher than [Fe {\sc VII}] 
disappeared and [Fe {\sc VII}] lines show a marginal decrease (Komossa et al. 2008; 
2009). In our MMT spectrum, only weak [Fe {\sc VII}] lines are detected. 

J1342+0530 --- its SDSS spectrum shows strong [Fe {\sc X}], [Fe {\sc XI}] and 
[Fe {\sc XIV}] but no [Fe {\sc VII}]. In the MMT spectrum, all [Fe {\sc X}], 
[Fe {\sc XI}] and [Fe {\sc XIV}] disappear but [Fe {\sc VII}] lines appear. 

J1350+2916 -- The variability of coronal line spectrum is similar to J1342+0530.

In summary, for all these sources, the coronal lines are shifted from high to low 
ionization species, or to a lack of coronal lines. 

\subsection{Other Narrow Emission Lines}

W12 showed that conventional narrow-line ratios measured from the SDSS spectra 
place these objects in the locus of star forming galaxies on the BPT diagrams 
(Baldwin, Phillips \& Terlervich 1981; also Kewley et al. 2006). This suggests 
that these lines come mainly from  star-forming regions around the nuclei. 
Figure 8 demonstrates the line ratios measured from the MMT spectra on the 
BPT diagrams. The line ratios are now very close to or above the line of demarcation 
between extreme AGNs and the star-formation region. This is at least partially 
attributed to the increase of [O {\sc III}] from the nucleus as discussed below.

%

Changes in the BPT diagram must be interpreted with caution because the SDSS and MMT 
spectra were obtained using different apertures. The MMT long-slit spectra are extracted 
from a smaller aperture (1"$\times$3") than the SDSS fiber size (3" diameter circle), and 
thus contain less light of the host galaxy. The emission lines from the extended star-formation region 
should be weaker in MMT spectrum than in SDSS spectrum. Indeed, the low ionization 
lines, including [S {\sc II}], [N {\sc II}], and narrow {\ha}, {\hb}, are substantially 
weaker in the MMT spectra than in the SDSS spectra(table 2), suggesting that they come from a large 
region. However, [O {\sc III}] in the MMT spectra is significantly stronger than in SDSS 
spectra (Figure 6 and table 2). The difference (360\%, 33\%,75\% and 50\% for J0748+4712, J0952+2143, 
J1342+0530 and J1350+2916) is larger than the calibration uncertainty (typically 4\% for 
SDSS and $<$5\% for MMT spectrum). This suggests that \oiii\ had brightened since the SDSS 
observation. The increment must come from the very nuclear region that is related to the 
tidal disruption event.

It is surprising that we measure an [O {\sc III}] doublet ratio \oiiis/\oiii$\simeq1/2$, 
which is very different from the conventional $1/3$, in J0748+4712 and J1342+0530 (Figure 6). 
To check for potential contamination from other lines, we examined the NIST atomic database for lines 
within 3\AA\ of \oiiis\footnote{Giving the similar profiles of [O {\sc III}] doublet in 
these two objects, contamination by lines with larger offset relative to \oiiis\ 
can be ruled out.}. Only Ti {\sc I}$\lambda$4958.3, Fe {\sc II}$\lambda$4958.2, Pm {\sc I}$\lambda$4959.5 
and Fe {\sc I}$\lambda$4961.2 are within 3\AA\ of {\oiiis}. Fe {\sc II} can be ruled out because 
it would produce much stronger emission at 4889.6\AA\ from the same upper 
level, which is not detected. For a similar reason, Fe {\sc I} can be excluded because it 
predicts a stronger emission line at 4916.3\AA. Ti {\sc I} and Pm {\sc I} have the same problem 
in addition to a very small element abundance. Therefore, contamination with other lines can be 
ruled out. We also looked for  potential telluric absorption lines and sky lines around 
the redshifted line wavelengths and found none. Since \oiiis\ and \oiii\ share the same upper level, 
the different line ratios can be only explained by the radiation transfer effect, which 
requires a substantial optical depth for the line. A detailed interpretation of this is given in 
\S \ref{sec:oiii}.

Narrow \heii\ shows a more complicated variability pattern. In J0952+2143, strong 
narrow \heii\ seen in the SDSS spectrum becomes barely detectable in the MMT spectrum, 
along with weakening of coronal lines and the broad lines. On the other hand, narrow 
\heii\ remains unchanged or shows a small increase in J1342+0530 in the course of 
the decrease of high ionization coronal lines and the appearance of \fevii. The situation 
is not clear in the other two objects because of either blending with the broad component 
in the SDSS spectrum or weakness of the line in the MMT spectrum. Note that \heii\ in 
both SDSS and MMT spectra are systematically broader than other low-ionization narrow 
emission lines, such as \ha\ and [S {\sc II}], which mainly come from star forming regions 
(Figure 7). This is consistent with the expectation that \heii\ is powered mainly by the 
continuum flare, and star-formation makes little contribution to it.

\section{Discussion\label{sec:discussion}}

\subsection{Formation and Evolution of Broad Lines}

In our sample, three targets (J0748+4712,J0952+2143,J1350+2916) show broad recombination 
lines ({\ha}, {\hb} or {\heii}) in the SDSS spectra and all these lines are double-peaked or 
strongly blue shouldered. Such lines can be formed in biconical outflows launched by the 
super-Eddington accretion flow, by a fast outflowing unbound stellar debris, or a ring of  
accreted debris, that are photoionized by strong radiation from the accretion disk. 
Bogdanovi{\'c} et al. (2004) showed that photoionized debris produces only weak broad 
\ha\ line ($L( {\rm H}\alpha)\sim 10^{37}$ erg~s$^{-1}$), which is much smaller than the observed 
one, for the tidal disruption of a solar type star by a black hole of $10^6$ M$_\sun$, 
because the stellar debris is confined within only a small subtending solid angle to 
the accretion disk. There are several possible way in which emission lines can be enhanced. 
First, if the star is evolved, e.g., a subgiant, the debris may spread out more in the vertical 
direction because an evolved star has a much large size, thus producing more reprocessed 
emission lines. Second, if the star's orbit mis-aligns with the spin direction of the black 
hole, the debris disk may receive more light from the accretion disk. Finally, the line 
equivalent width is likely also dependent on the mass of black hole. Small black holes 
may produce larger line equivalent widths because the tidal disruption radius is smaller 
so the stellar debris has a larger covering factor. It remains to be seen whether taking into 
account all these effects can reproduce the observed line equivalent width. Thus we favor
the biconical outflow model because it predicts a large line equivalent 
width and double-peak profile which are more consistent with observed lines (Strubbe \& Quataert 2009).

In either scheme, broad lines are expected to fade on the time scales of months to a 
year. The fading is caused by both the declining of the ionizing continuum and the 
evolution of the debris properties or outflows. The variation 
of the ionizing continuum is qualitatively understood, a nearly constant luminosity 
at the super-Eddington phase and a power-law decay phase with an index around 
$-5/3$ to $-5/2$ (Rees 1998; Laodato \& King 2009). In either stage, the ionizing 
continuum becomes softer. In a quasi-steady accretion system, it is expected that the 
outflow weakens as the accretion rate decreases. Since the accretion rate varies 
on dynamic time scales, the structure of the outflow is likely complicated. The disk 
initially launches strong outflows during the super-Eddington phase, which is accelerated 
to a high velocity.  As the accretion rate decreases, the disk will produce a weak 
outflow, and the radiation acceleration is greatly reduced; after certain time, 
it may eventually become smaller than the gravitational one, and then outflow 
decelerates in the gravitational field.   

In J0748+4712 and J1350+2916, we can only put  upper limits on the lifetime of
the broad lines to 5 years. In J0952+2143, we witness the declining of the broad lines. In 
the NTT spectrum taken 768 days after SDSS spectrum, the redshifted broad Balmer 
emission lines and double horn profile are still prominent (Komossa et al. 2009). 
The line flux declined by a factor of only 2.8 in comparison with SDSS observation. 
This is slower than a $\propto t^{-4/3}$ or $t^{-5/3}$ law if we take the initial flare to have occurred 
near the SDSS photometric observation, 375 days before the SDSS spectroscopic observation. 
In its MMT spectrum, which was taken 2187 days after the SDSS spectrum, the double horn 
profile disappeared but the weak redshifted broad component still exists. Taking into 
account the short recombination time scale of excited H atoms ($\approx 10^5/n_e$ 
yr, Osterbrock \& Ferland 2006), an  ionizing continuum is needed to produce 
the broad Balmer lines. The continuum may form through late stage accretion of residuary 
stellar debris in very eccentric orbits, or by fallback of failed outflows. 
The broad line flux is a factor of 4 lower than in the NTT spectrum. Noting that both the 
line width and the centroid offset decreases with time, this is consistent with the decelerating 
outflow scenario.  

Probably due to the relative small size of the accretion disk, outflows from both 
sides are observed, leaving a double peaked profile, in contrast to the single-peak 
blueshifted profiles of CIV in luminous quasars (Richards et al. 2011; Wang et al. 
2011). The He {\sc II} line in J0748+4712 is not symmetrical around zero velocity 
at the source rest frame. This perhaps can be attributed to the partially obscuration 
of the base of the receding outflow. The small separation between the two peaks in the other 
two objects can be produced by either a large inclination to the collimated outflows or a small outflow 
velocity. As noted in W12, J0748+4712 was seen by SDSS much earlier than the other two 
objects when the disk emission is still strong and outflows show a high velocity. With 
the decrease of the accretion power, the outflows are decelerated in the gravitational 
potential of the black hole, leading to small outflow velocities in other two objects.  

As discussed in W11 (see also Peterson \& Ferland 1986; Gezari et al. 2012), the large 
He {\sc II}/{\ha} ratio requires an over-abundance of helium relative to hydrogen, which 
is interpreted by W11 as the tidal disruption of an evolved star. Following Peterson \& Ferland (1986) and assuming a gas temperature of $2\times10^4K$, a minimum of $n_{He}/n_{H}=0.75$ is required by assuming that most of helium is in He$^{+2}$.

Most TDEs discovered by X-ray and UV flares do not show broad emission lines, while three 
of four objects in this sample do. This may be attributed to selection effects. On the one 
hand, the broad line objects are usually considered as AGN and rejected for further 
monitoring to reduce the number of sources for repeated spectroscopic follow-up (Gezari et al 
2009; van Velzen et al 2011; Cenko et al 2012). On the other hand, coronal-line selected 
objects preferably lie in the gas-rich disk galaxies, and they tend to host small black 
holes. The accretion rate is expected to be higher than that of a high- mass black hole, 
and outflows should be denser and stronger. If broad lines are formed in outflows, one 
naturally expect that TDE by small black holes produce strong broad emission lines. More 
theoretical calculations are required to verify this. 

\subsection{Consequences of optically thick [O {\sc III}] emission\label{sec:oiii}}
\subsubsection{Simple estimates}
The [O {\sc III}] line ratio, which is $\sim 2$ rather than the expected 3, suggests that the lines are optically thick.
In this section we examine some consequences of this observation.
This discussion is preliminary because there is not yet sufficient spectroscopic constraints
to fully define a model for the observations.

If a line is thermalized, that is, the density is substantially above the critical density
of the line, then the level populations are given by the gas kinetic temperature
and line photons are lost in the scattering process because they are collisionally deexcited. 
The intensity of a line is then given by
$I_{\nu} = B_{\nu} [1-\exp(-\tau) ]$ where $B_{\nu}$ is the Planck function and $\tau$ the line optical depth.
The $\lambda\lambda$5007, 4959 line optical depths $\tau$ are in a 3:1 ratio, so the intensity ratio is given by
$I(\lambda 5007)/I(\lambda 4959)= [1-\exp(-\tau) ]/[1-\exp(-\tau/3) ]\approx 2$.
This function is shown in Figure 9.
A line ratio of $\approx 2$ corresponds to an optical depth $\tau(\lambda 5007) \approx 1.5$.

The actual line ratio is a function of both optical depth and density.  
If the electron density is well below the critical density of the $^1$D$_2$ level that produces the [O {\sc III}] lines
then photons will be remitted after absorption. They will simply scatter out of the cloud with no
loss of intensity and the 3:1 line ratio is maintained.
Line photons are only lost when there is a large probability of collisional deexcitation, 
they are thermalized, following absorption.
The lines will be well thermalized when the density is
$n_e \gg n_{crit}(^1{\rm D}_2) \approx 6.8\times 10^5\ t_4^{0.5}\ {\rm cm}^{-3}$,
where $t_4 = T/10^4$~K.
If the density is less than this, a larger $\lambda 5007$ optical depth would be needed to reach the same line ratio.

More O {\sc III} lines would be needed to estimate the parameters for the [O {\sc III}]-forming region.
A good detection of the $\lambda 4363$ line would be important.
These observations are not available so 
we will simply assume that the $\lambda 5007$ optical depth is of order unity, $t_4 = 10^4$~K,
and that the electron density is of order the $^1$D$_2$ critical density.

\subsubsection{Column density and Compton depth}

The line absorption coefficient is given by 
\begin{equation}
\alpha = 1.497\times 10^{-6} f_{lu} \lambda_{\micron} / u_{Dop}\ {\rm cm}^2
\end{equation}
where $f_{lu}$ is the absorption oscillator strength, $\lambda_{\micron}$ the wavelength in microns,
and $u_{Dop}$ the Doppler width in velocity units.
Assuming the transition probabilities given by Storey \& Zeippen(2000)
we find $\alpha = 5.57\times 10^{-17} u_{Dop}^{-1}\ {\rm cm}^2$.

The Doppler width is given by the thermal width if turbulence is absent;
\begin{equation}
\label{eq:LineWidth}
u_{th}= \sqrt {2kT/m} \ {\rm cm\ s}^{-1} \approx 12.8 \sqrt{ t_4/m_{AMU}}\ {\rm km\ s}^{-1}
\approx 3.2\ t_4^{1/2} \ {\rm km\ s}^{-1}
\end{equation}
where the expression is evaluated for oxygen, $m_{AMU} = 16$.
The absorption cross section becomes
\begin{equation}
\alpha = 1.74 \times 10^{-22} \ t_4^{-1/2} \ {\rm cm}^2
\end{equation}

The column density required for $\tau(\lambda 5007) > 1$ is then
\begin{equation}
N({\rm O}^{2+}) > 5.75\times 10^{21} \ t_4^{1/2} \ {\rm cm}^{-2}
\end{equation}
The solar O/H ratio is $4.9\times 10^{-4}$ (Asplund et al.2009) so the total hydrogen 
column density at a metallicity $Z$ will be
\begin{equation}
\label{eq:HColDen}
N({\rm H}) > 1.2\times 10^{25} \ t_4^{1/2} \ Z^{-1}\ {\rm cm}^{-2}
\end{equation}
The corresponding electron scattering optical depth is then
\begin{equation}
\tau(e) = \sigma_T N({\rm H}) > 7.8 \ t_4^{1/2} \ Z^{-1}
\end{equation}
where $\sigma_T$ is the Thomson cross section.
This has the important consequence that, unless the metallicity is quite large ($Z\ge 10$),
the [O {\sc III}] line-forming region is optically thick to electron scattering.

This means that we may have underestimated the line width in Equation \ref{eq:LineWidth}
since $\lambda 5007$ photons will scatter off the rapidly-moving electrons.
The mass of an electron is $2.94\times 10^4$ smaller than an oxygen atom so the line
width is 171 times wider. The column densities are increased by $\sqrt{171}\simeq13$ 
if electron scattering dominates the line width.

The thermal width of [O {\sc III}] $\lambda 5007$ at $10^4$~K would be $\sim 550$~km~s$^{-1}$ 
in the electron-scattering dominated limit. This is substantially larger than the observed 
widths, 200 and 255 km~s$^{-1}$ \footnote{This width is $\sqrt{2} \sigma_v$.}, showing that 
the lines are not broadened into the full electron scattering limit. At optical depth 
$\tau(e)<1$, the scattered photons will produce an extended wing with a width dependent on 
the optical depth and electron temperature, supposed on the primary emission profile (Laor 
2006). Thus the profile of high velocity wing can be used to constrain the properties of 
the scatter. Unfortunately, from our data we can not 
confirm or rule out the existence of the broad wing of [O {\sc III}] emission lines. The lack of a 
prominent wing would indicate that the gas has a high metallicity of order $Z\sim10$. 

\subsubsection{Suggestions from photoionization models}

The large column density suggested in Equation \ref{eq:HColDen} places constraints
on the radiation field shape and intensity.
In its simplest form the photoionization balance equation may be written as (Osterbrock 
\& Ferland(2006))
\begin{equation}
\phi({\rm H}) = n_e \, n_p \, \alpha_B \, dl
= n_e \, \alpha_B \, N({\rm H})
\end{equation}
where $\alpha_B$ is the hydrogen Case B recombination coefficient, 
$\phi({\rm H})$ is the flux of hydrogen-ionizing photons, and $dl$ is the
Str\"omgren thickness, the thickness of the H$^+$ layer.
This can be written in terms of the dimensionless ionization parameter $U$,
the ratio of densities of hydrogen-ionizing photons to hydrogen,
\begin{equation}
U({\rm H}) \equiv \frac{\phi({\rm H})}{n({\rm H}) c} = 1.1\, \frac{\alpha_B}{c} \, N({\rm H})
\end{equation}
where we assume that helium is single ionized with solar metallicity, so that $n_e = 1.1n({\rm H})$.
Putting in numerical values and assuming
$\alpha_B \approx 2.6\times 10^{-13}\, t_4^{-0.8}$, the limit in Equation \ref{eq:HColDen} becomes
\begin{equation}
U({\rm H}) > 10^2\, t_4^{-0.3}\, Z^{-1}.
\end{equation}

This is a large ionization parameter.
Photoionization models of strong-[O {\sc III}] lined objects are often fitted with
$U \sim 10^{-2} - 10^{-1}$.
For photoionization by a continuum that extends to high energies, such as
the SED of an AGN, the ionization of the gas is proportional to $U$ (Osterbrock 
\& Ferland 2006) and 
little O$^{2+}$ is present for such large values.

This is not the case if the SED is soft. If few ionizing photons are present
at energies that can ionize O$^{2+}$ to O$^{3+}$ then the required column density of O$^{2+}$
will be produced.
This occurs if the SED is equivalent to a blackbody with $T < 5\times 10^4$~K, or
if a more energetic SED is filtered through in intervening absorbing column which
removes high-energy photons.
This brings in the nature of the ionizing source, a fundamental question in these objects.

A very large $U$ with a soft SED is consistent with the 10$^4$~K temperature we have
assumed. In photoionization equilibrium the gas heating is proportional to the photoionziation
rate, which is equal to the recombination rate. There is no direct dependence on 
the flux of photons or the ionization parameter.
A stronger flux of photons produces higher ionization but the same photoionization rate,
which is equal to the recombination rate, so the heating rate is constant.

\subsection{Formation and Evolution of Narrow Emission Lines}

Coronal lines and a significant fraction of [O {\sc III}] must come from ambient gas 
photoionized by the flare as they are variable on time scales of several years. Most 
previous TDE candidates that were discovered via X-ray or UV flares do not have such 
features. The recent discovered TDE candidate PS1−10jh with a strong broad \heii\ line but weak Balmer lines as J0748+4712 do not show coronal lines either (Gezari et al. 2012).
Thus only a fraction of TDE shows coronal lines, and the presence of coronal 
lines can be independent of the existence of broad lines. This can be understood 
in the framework already discussed. The presence of coronal lines is related to the gas 
distribution close to the supermassive black hole (W12), while 
the formation of broad lines depends on the strength of outflows or the geometry of 
debris relatively to the disk radiation, which is likely anisotropic. Although the nuclear 
gas environment and outflows are expected to be correlated with the black hole mass in 
the local universe statistically, there are many outliers. 

Komossa et al. (2008) postulated that coronal lines are echoes of the continuum flare 
by distant gas. As discussed in W12, the variations of emission lines are rather complex 
in this scenario, depending on the time evolution of the flare as well as the distribution 
of gas surrounding the massive black hole. The emission lines at a given time come from 
different regions that are illuminated by the continuum with different time advances. Thus, in 
the optically thin case, the optimal emission region for a specific line is determined by 
competition between the time fading of the continuum, the position dependent time lag, the $r^{-2}$ 
dilution of incident continuum intensity, and the radial dependence of density. Figure 10 
shows the snapshot of ionization parameter distribution for a model in which the ionizing 
continuum varies with time as a power-law declining phase $L(t)\propto t^{-5/3}$, and a 
smooth density distribution $n \propto r^{-\beta}$. In the optically thick case, additional 
frequency dependent attenuation has to be considered, which steepens the ionizing continuum 
at distant region. 

Even with the sparse sampling, the observations clearly require a different duration 
time scale of coronal lines in different objects. Coronal lines in J1342+0530 lasted for 
a time scale of at least ten years with a transition from high ionization coronal lines 
only to low ionization lines only, while in J0748+4712, all coronal lines disappeared in 
less than 4-5 years. It is not clear what causes the different time scale: the continuum 
light curve or the gas distribution. We do not know what happened between the SDSS observation 
in 2004 and Xinglong 2.16m observation in 2009 for J0748+4712. Was there a similar transition 
from high ionization coronal lines only to low ionization coronal lines, or all coronal lines 
disappeared simultaneously? Even more, also for J1342+0530 and J1350+2916, is there an 
intermediate state with both high and low ionization coronal lines as seen J0952+2143? If so, 
then there is a continuous transition from high to low ionization state and we can unify the 
high and low ionization coronal line emitter through time evolution. Unfortunately, we do not 
have the data in these important evolution stages. 

[O {\sc III}] variability provides additional constraints on the gas distribution in 
the inner parsecs and the evolution of the ionizing continuum. The brightening of [O {\sc III}] 
follows the declining of coronal lines, i.e., average ionization of line emitting gas 
decreases continuously. Giving the short recombination times of O$^{2+}$ ($1.3\times
10^5\ (10^6\ {\rm cm}^{-3}/n_e)$~s), the gas is in quasi-ionization equilibrium. This 
requires either attenuation of the ionizing continuum or/and the density must decrease more 
slowly than $r^{-2}$. The analysis in the last section suggests a soft ionizing continuum. 
Combining with the strong coronal line emission in the early spectra, the supports the idea that hard 
ionizing photons had already been filtered. 

We have derived a lower limit on the gas column density in the last section for two 
sources (J0748+4712 and J1342+0530) with non-canonical [O {\sc III}] ratios.  
Bearing in mind that the flare is less than ten years, so $O^{2+}$ ions exist within 
the region bounded by the parabolic surface with a lag of ten years (see Figure in W12).
The large column density sets a lower limit on the gas density of order $n({\rm H})\sim
1.2\times 10^{25} \ t_4^{1/2} \ Z^{-1}/c\ \Delta t \ {\rm cm}^{-3}=2\times 10^6 \ t_4^{1/2} 
\ Z^{-1} \ \Delta t_6^{-1} \ {\rm cm}^{-3}$, where $\Delta t_6=(\Delta t/6)\ {\rm yr}$.

Assuming that all observed [O {\sc III}] comes from the same region, we can derive 
the volume and mass of [O {\sc III}] emitting gas. At a temperature $10^4$ K, the volume 
emissivities of \oiii\ are 2.0$\times 10^{-16}$, 1.7$\times 10^{-14}$, 6.7$\times 10^{-13}$ 
and 9.5$\times10^{-12}$ erg~s$^{-1}$~cm$^{-3}$ for $n({\rm H})=10^4$, $10^5$, $10^6$ and 
$10^7$ cm$^{-3}$ assuming a solar abundance. The observed \oiii\ luminosities for J0748+4712 
and J1342+0530 are similar and around 9$\times10^{39}$ erg~s$^{-1}$, requiring a volume of 
line emitting gas of only 2.7$\times10^{52}$ cm$^3$ for $n({\rm H})=10^6$ cm$^{-3}$. This volume 
is only a small fraction of the volume swept by the flare radiation in nearly 10 years, suggesting 
a small filling factor. Combining this with a large column 
density implies that the emission region consists of only a small number of thick clouds.

  

\subsection{Could the Powering Source be AGN Variability?}
A substantial fraction of AGNs show coronal lines, but their strengths are usually 
much weaker than [O {\sc III}]. In the SDSS spectra, coronal lines of these TDE candidates 
are 1-2 orders of magnitude stronger than that of strong coronal line emitting Seyfert galaxies. Further,
large body of AGN monitoring have shown that coronal lines in most 
AGNs are quite stable, while in rare cases which coronal lines do vary on time scale 
of years and the amplification is usually moderate (Penston et al.1984; Veilleux 
1988). Only in one extreme case IC 3599, a strong decline of [Fe {\sc X}] by two orders of 
magnitudes was observed after an unusual soft X-ray burst by a factor of 100, that 
was sometimes interpreted as tidal disruption event (Brandt et al. 1995; Grupe et al. 
1995; Komossa \& Bade 1999). In our four TDE targets, coronal lines with ionization 
potentials higher than that of [Fe {\sc VII}] show strong fading of more than two orders of 
magnitudes within ten years and [Fe {\sc VII}] lines appear in two targets without them in 
SDSS spectra. That makes them very extreme if they were AGNs. In addition, large amplitude variations (a factor of 100) in the highly ionized coronal lines require large amplitude variations in soft X-rays. Up to now only narrow line Seyfert 1 galaxies and BL Lac objects are known to show such large amplitude variability in X-rays (Eracleous et al.2012). All the seven ECLEs targets are covered by FIRST survey (Becker et al. 1995), but only J0938+1353 was detected with a low flux of 1.93 mJy, while others are below the detection limit, typical 1 mJy. So from optical spectrum and radio flux, we can rule out both possibilities.

Along with the variation of coronal lines, a significant increase of [O {\sc III}] emission lines is detected in 
the four TDE targets. There is no clear evidence that emission lines such as [O {\sc III}] 
from normal AGN NLR would show strong variability due to its large size. So persistent AGN activity is very unlikely the powering source of the variation of coronal lines and [O {\sc III}] lines.

The disappearance of broad emission lines are also very different from other Seyfert 
galaxies monitoring so far. In these Seyfert galaxies, broad emission lines usually 
vary in response to changes in the continuum with a moderate amplitude. Only two cases were reported for temporary disappearance of broad emission lines over a period of a few months accompanying with large continuum variations in NGC 4151 and NGC 5548 (Penston \& Perez, 1984; Iijima et al. 1992). 
In J0748+4712 and J0952+2143, which have several observations after the outburst, we only find broad 
lines are either fading or disappeared. Therefore, there is no evidence for re-starting of the central engine.
Considering all these facts, we believe that the powering source is likely a strong 
outburst in an quiescent galaxy rather than stochastic variability of an AGN.

\subsection{Could the Powering Source be Supernova Explosion?}
For the Supernova (SN) scenario, it is unable to explain the high luminosities of coronal lines and the absence of other low-ionization lines, and the continuum variability in some targets(W12).
Also, the MMT follow-up observations provide additional constraints on the SN 
scenario. First, four TDE candidates show an increase of 
[O {\sc III}] between SDSS to MMT observations, and the [O {\sc III}] are 
systematically broader than normal low-ionization lines such as {\ha} (figure 7). 
This characteristic is different from other known SN. Very few young SNs display 
variable narrow [O {\sc III}] emission. 
Some SNs do show [O {\sc III}] emission a few years to up 100 years after 
explosion, but these lines are usually broad, accompanying low ionization 
lines of similar widths (Milisavljevic et al. 2012). They are formed 
in the SN shell interacting with the stellar winds. 
Second, based on the total energy in the high ionization coronal lines, W12 estimated 
a total energy in soft X-ray of order $10^{50}$ ergs. The emergence of 
[Fe {\sc VII}] lines in these two objects with only high ionization coronal lines 
in previous SDSS spectra (J1342+0530 \& J1350+2916) and the persistence of [Fe 
{\sc VII}] in J0952+2143 suggest that a strong UVX ionizing continuum is still 
seen by the line emitting gas. The brightening of the {\heii} line in J1342+0530 indicates 
a further increase of EUV ionizing photons absorbed by line emission gas in this object, 
probably arising from large covering factor of gas at the time lag. These variations 
of [Fe {\sc VII}] and {\heii} lines re-enforce the conclusion in W12 that a very 
energetic flare is required to power the coronal line emission. 

As shown in the last section, the presence of an optically thick [O {\sc III}] emission region 
requires large ionization parameters. In echo models, gas is photoionized by the continuum 
flare that took place 5-9 years ago, so the size of emission line region is $R_{\rm [OIII]}
\simeq 1-3$ pc. The peak luminosity of ionizing continuum can then be estimated from the 
ionization parameter, the distance and the density,
\begin{equation}
L_{ion}=4\pi R_{\rm [O \sc III]}^2 n({\rm H})U({\rm H})c \langle h\nu \rangle > 7.9\times 10^{45} t_4^{-0.3}Z^{-1}R_{pc}n_6({\rm H}) \ {\rm erg\ s}^{-1}
\end{equation}
with a reasonable $n_6({\rm H})=n({\rm H})/10^6 {\rm cm}^{-3}\sim 1$, the 
luminosity would be $10^{45-46}$ erg s$^{-1}$, which is much more luminous than any supernova and is
close to the Eddington limit for a $10^{7-8}$ M$_\sun$ black hole.  
Finally, the life time of broad lines are at most several years, which is much shorter 
than these in type II SN. 


\section{Conclusions}

We have carried out follow-up MMT observations of seven extreme coronal line emitters 
that were studied by W12. Three objects turn out to be persistently strong coronal line emitters. Among them, 
two are likely star-forming and AGN composite or tidal disruption of a giant star, and 
one is a narrow line Seyfert 1 galaxy. In the other four objects, all coronal lines with 
ionization higher than [Fe {\sc VII}] disappear. Thus, they are tidal disruption candidates. 
[Fe {\sc VII}] lines faded in J0952+2143 and appear in the MMT spectra of two objects with higher 
ionization coronal line only in SDSS spectra. [O {\sc III}] doublets are brightened in all 
four objects. Thus we are still witnessing echoes of the continuous decrease of the 
gas ionization in the emitting region. If this trend continues, the object will appear in the 
AGN locus of the BPT diagram. Further monitoring of the emission lines is needed. 

We also detected non-canonical [O {\sc III}] ratios in two objects. The column density of $O^{2+}$ 
must be large to make [O {\sc III}]5007 optically thick. For reasonable metallicity, the H column 
density is large and the gas is probably optically thick to electron scattering. Lines will be 
broadened due to electron scattering, which can be tested with future high quality data. The ionization 
parameter has to be very large to get this column density, and this also requires that the ionizing 
SED be relatively soft, equivalent to a $T < 5\times 10^4$~K blackbody, or oxygen would be too 
ionized. A gas temperature of order 10$^4$~K will occur for such extreme conditions due to the
nature of photoionization equilibrium, if the SED is this soft.

These observations can be fitted into a picture where giant molecular clouds are illuminated by  
continuum flares with an ionizing continuum luminosity of order $10^{45-46}$ erg~s$^{-1}$ within 
a few parsecs of the nucleus. The gas in the inner face of the cloud produces the coronal lines. 
As high energy photons are absorbed by the gas, the ionizing continuum illuminating outer part of 
the clouds softens and creates a thick [O {\sc III}] emission region. 
The presence of giant molecular clouds in the center parsec and 
no persistent nuclear activity is striking, suggesting that perturbation in the gas is 
more fundamental for the black hole fueling. Clearly, we still miss some critical evolution 
phases, which prevents us from probing a direct connection between these two subclasses of  extreme 
coronal line emitters. Future follow-ups of new discoveries of such objects from the 
spectroscopic surveys of LAMOST and BigBOSS can fill the gaps and yield a more complete 
picture of emission line evolution.

Broad emission lines previously seen in the SDSS spectra of J0748+4712 and J1342+0530 vanished 
in the MMT spectra, and there is only weak broad Balmer lines in J0952+2143. In J0952+2143, we 
observed continuously fading of the broad lines by a factor of 11 in past 8 years as well 
as a decrease of the line width and velocity shift. The line profile variability perhaps 
reflects the deceleration of emission line gas in the gravitational field of the black hole 
due to a lack of further radiative acceleration. With improved subtraction of stellar light, 
the broad lines in these objects are found to all be double-peaked or blue-shouldered, suggesting it 
is a general property, probably from biconical outflows. 

\acknowledgements
We would like to thank Cai Zheng, Zhen-Ya Zheng and Hui Dong for their great help in MMT observation 
and Stefanie Komossa for her very useful discussion. 
This work is supported by NSFC 11233002 and 10973013. This research uses data obtained
through the Telescope Access Program (TAP), which is funded by the National Astronomical 
Observatories, Chinese Academy of Sciences, and the Special Fund for Astronomy from the 
Ministry of Finance. Observations reported
here were obtained at the
MMT Observatory, a joint
facility of the University of
Arizona and the Smithsonian 
Institution.
GJF acknowledges support by NSF (1108928; and 1109061),  and STScI (HST-AR-12125.01, GO-12560, and HST-GO-12309).

\begin{deluxetable}{ccccccccccc}
\tabletypesize{\scriptsize}
\rotate
\tablecaption{Basic Data of Extreme Coronal Line Emitters\tablenotemark{a}} \label{tab1}
\tablewidth{0pt}
\tablehead{
\colhead{No} & \colhead{Name} & \colhead{$z$} & \colhead{$M_{i,{\rm tot}}$\tablenotemark{b}}& \colhead{Obs Interval\tablenotemark{c}} & \colhead{Broad Line} & \multicolumn{2}{c}{[Fe {\sc VII}]} & \colhead{[O \sc{III}]}& \multicolumn{2}{c}{NSC\tablenotemark{d}} \\
\cline{7-8} \cline{10-11}\\
\colhead{} & \colhead{} & \colhead{} & \colhead{}& \colhead{}& \colhead{} & \colhead{SDSS}   & \colhead{MMT} & \colhead{}  & \colhead{SDSS}   & \colhead{MMT}
} 
\startdata
1& SDSS J074820.66+471214.6 & 0.0615 & -19.75 & 2866 & fading  & no & no  & increase  & yes & no \\
2& SDSS J095209.56+214313.3 & 0.0789 & -20.41 & 2187 & fading  & yes& yes & increase  & no & no \\ 
3& SDSS J134244.42+053056.1 & 0.0366 & -18.91 & 3548 & absence & no & yes & increase  & no & no \\
4& SDSS J135001.49+291609.7 & 0.0777 & -19.76 & 2073 & fading  & no & yes & increase  & no & no \\

5& SDSS J093801.64+135317.0 & 0.1006 & -21.29 & 1834 &      &    &     &     &    &     \\
6& SDSS J105526.43+563713.3 & 0.0743 & -20.01 & 3548 &      &    &     &     &    &     \\
7& SDSS J124134.26+442639.2 & 0.0419 & -19.95 & 2859 &      &    &     &     &    &     \\
\enddata
\tablenotetext{a}{First four objects show significant variations in continuum and emission lines between two boservation, last three do not. We just list the variations of the first four objects.}
\tablenotetext{b}{$M_{i,{\rm tot}}$ are estimated from SDSS photometry.}
\tablenotetext{c}{Days bewteen SDSS and MMT observation.}
\tablenotetext{d}{NSC stands for Non-stellar conitnuum.}
\end{deluxetable}

\begin{deluxetable}{cccccccccccccccc}
\tabletypesize{\scriptsize}
\rotate
\tablewidth{0pt}
\tablecaption{Narrow Emission Line Flux$\tablenotemark{a}$ of Variable Targets} \label{tab2}
\tablehead{
\multicolumn{1}{c}{No} & \multicolumn{1}{c}{\ha$\tablenotemark{n}$} & \multicolumn{1}{c}{\hb$\tablenotemark{n}$} & \multicolumn{1}{c}{\ha$\tablenotemark{b}$} & \multicolumn{1}{c}{\hb$\tablenotemark{b}$} & \multicolumn{1}{c}{[Ne {\sc III}]} & \multicolumn{1}{c}{[He {\sc II]}} & \multicolumn{1}{c}{[N {\sc II}]} & \multicolumn{1}{c}{[O {\sc I}]} & \multicolumn{1}{c}{[O {\sc III}]} & \multicolumn{1}{c}{[O {\sc III}]} & \multicolumn{1}{c}{[O {\sc III}]} & \multicolumn{1}{c}{[Fe {\sc VII}]} & \multicolumn{1}{c}{[Fe {\sc VII}]} & \multicolumn{1}{c}{[S {\sc II}]} & \multicolumn{1}{c}{[S {\sc II}]} \\
\colhead{} & \colhead{} & \colhead{} & \colhead{} & \colhead{} & \colhead{$\lambda$3896} & \colhead{$\lambda$4686} & \colhead{$\lambda$6583} & \colhead{$\lambda$6300} & \colhead{$\lambda$4363}& \colhead{$\lambda$4959} & \colhead{$\lambda$5007} & \colhead{$\lambda$5722} & \colhead{$\lambda$6088} & \colhead{$\lambda$6716} & \colhead{$\lambda$6731}
}
\startdata
1& $131(4)$ & $49(4)$ & & & $<2$ & $<6$ & $48(4)$ & $12(3)$ & $<8$ & $31(4)$ & $65(4)$ & $<3$ & $<4$ & $23(4)$ & $21(4)$ \\
2& $108(7)$ & $26(4)$ & $155(10)$ & $39(7)$ & $31(4)$ & $11(3)$ & $64(5)$ & $14(4)$ & $<16$ & $72(3)$ & $215(3)$ & $13(3)$ & $17(5)$ & $25(4)$ & $27(4)$ \\
3& $171(5)$ & $27(4)$ & & & & $33(4)$ & $49(4)$ & $20(5)$ & $<15$ & $92(3)$ & $182(3)$ & $18(3)$ & $31(4)$ & $23(5)$ & $17(5)$ \\
4& $79(5)$ & $15(3)$ & & & & $6(2)$ & $22(4)$ & $<5$ & $<9$ & $22(1)$ & $66(1)$ & $8(2)$ & $12(4)$ & $9(2)$ & $9(2)$  \\
\enddata
\tablenotetext{a}{Emission line fluxes are in units of $10^{-17}$ erg~cm$^{-2}$~s$^{-1}$. A superscript 'n' means narrow line compoments and a superscript 'b' means broad line compoments.}
\end{deluxetable}

\begin{figure}
\epsscale{1.0}
\plotone{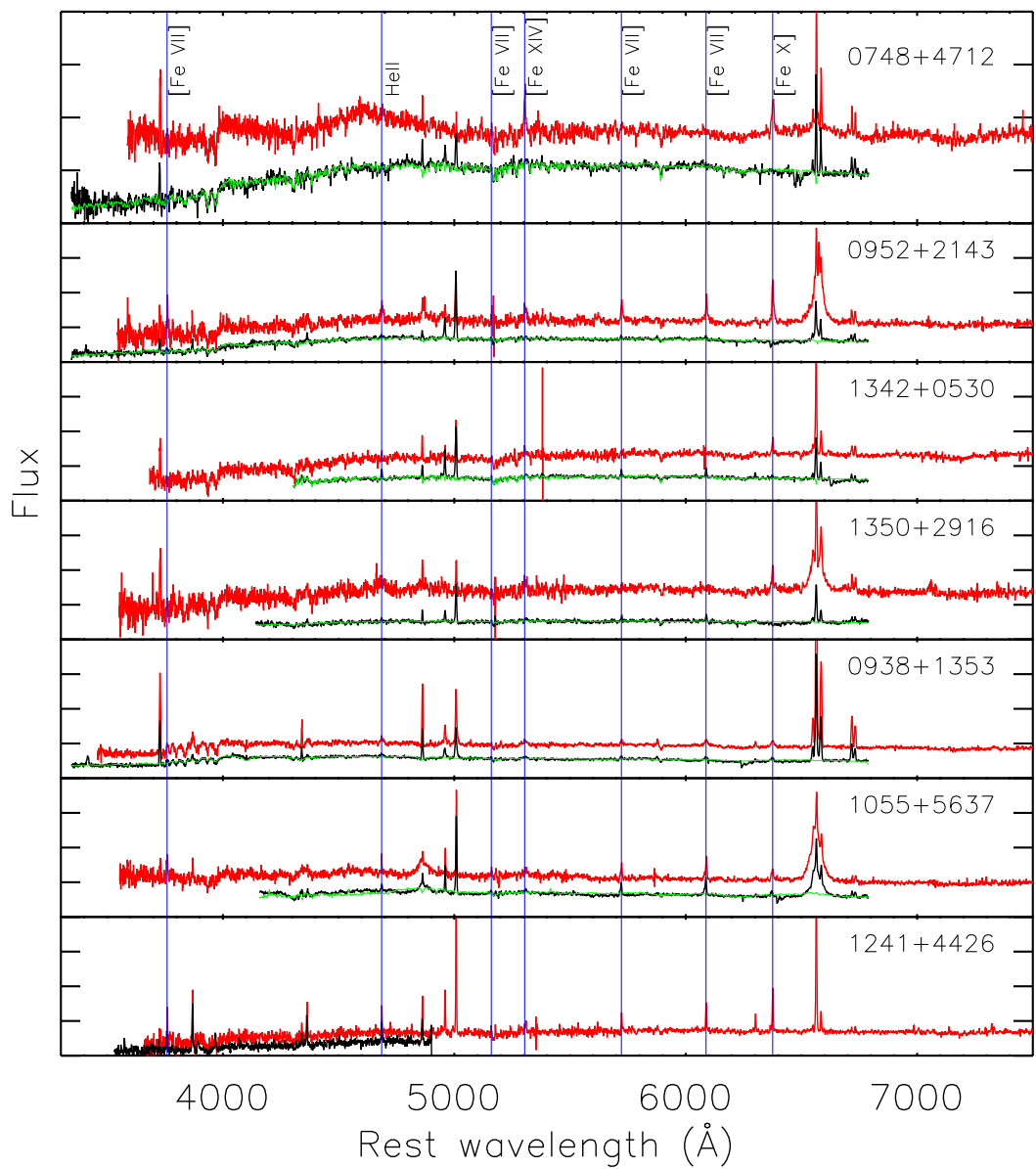}
\caption{MMT(black) and SDSS(red) spectra of seven extreme coronal line emitters from W12. Continuum fit are plotted in green. Coronal lines and \heii\ are marked in blue. The spectra have been shifted in vertical direction for clarity.}
\label{fig1}
\end{figure}     

\begin{figure}
\epsscale{1.0}
\plotone{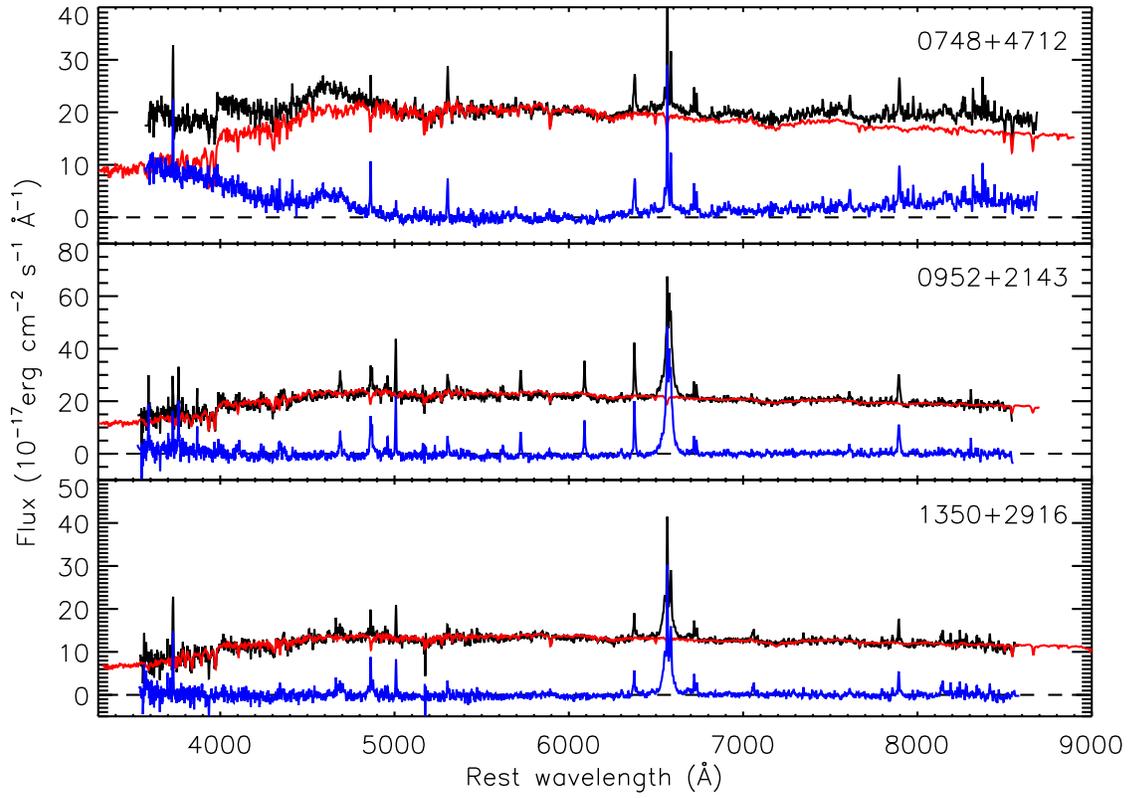}
\caption{SDSS spectra of J0748+4712, J0952+2143, J1350+2916 (black) and scaled up MMT spectra's 
6ICs model (red) and the residuals of the substraction (blue). Both the spectra and models 
are smoothed by 3 pixels.}
\label{fig2}
\end{figure}  

\begin{figure}
\epsscale{1.0}
\plotone{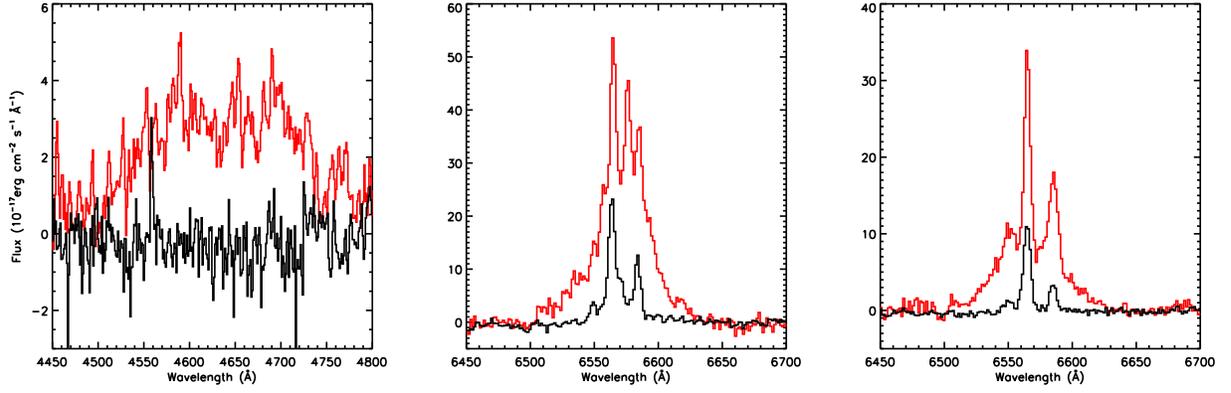}
\caption{Broad bump around 4600\AA\ in J0748+4712 (left panle,smoothed by 3 pixals) and \ha\ regions of J0952+2143 (middle panel) and J1350+2916 (right panel). MMT spectra are plotted in black 
and SDSS spectra are plotted in red, all the spectra have been continuum substracted.}
\label{fig3}
\end{figure}

\begin{figure}
\epsscale{1.0}
\plotone{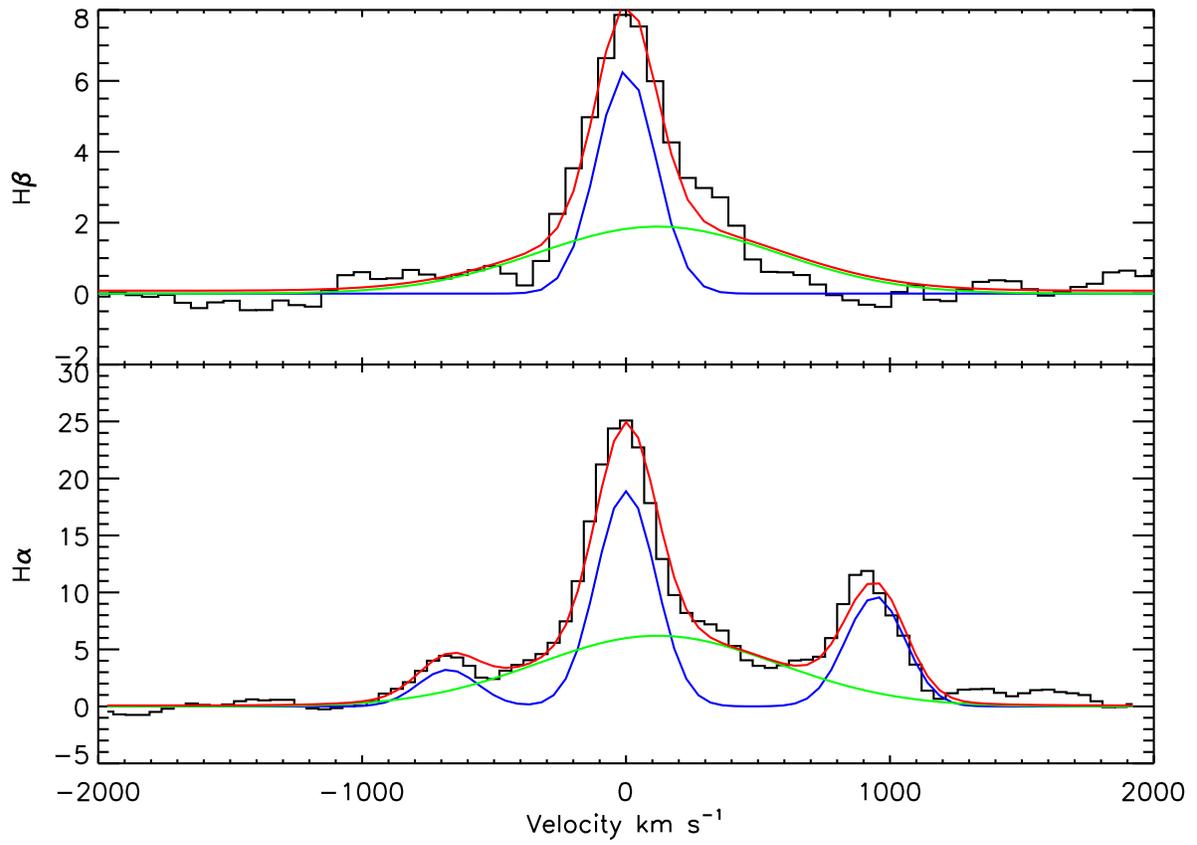}
\caption{J0952+2143's \ha\ (low pannel) and \hb\ (up pannel) line profiles(black) in MMT spectrum. Normal narrow component fit is ploted in blue, redshift broad component is ploted in green. Model sum is plotted in red.}
\label{fig4}
\end{figure}  

\begin{figure}
\epsscale{1.0}
\plotone{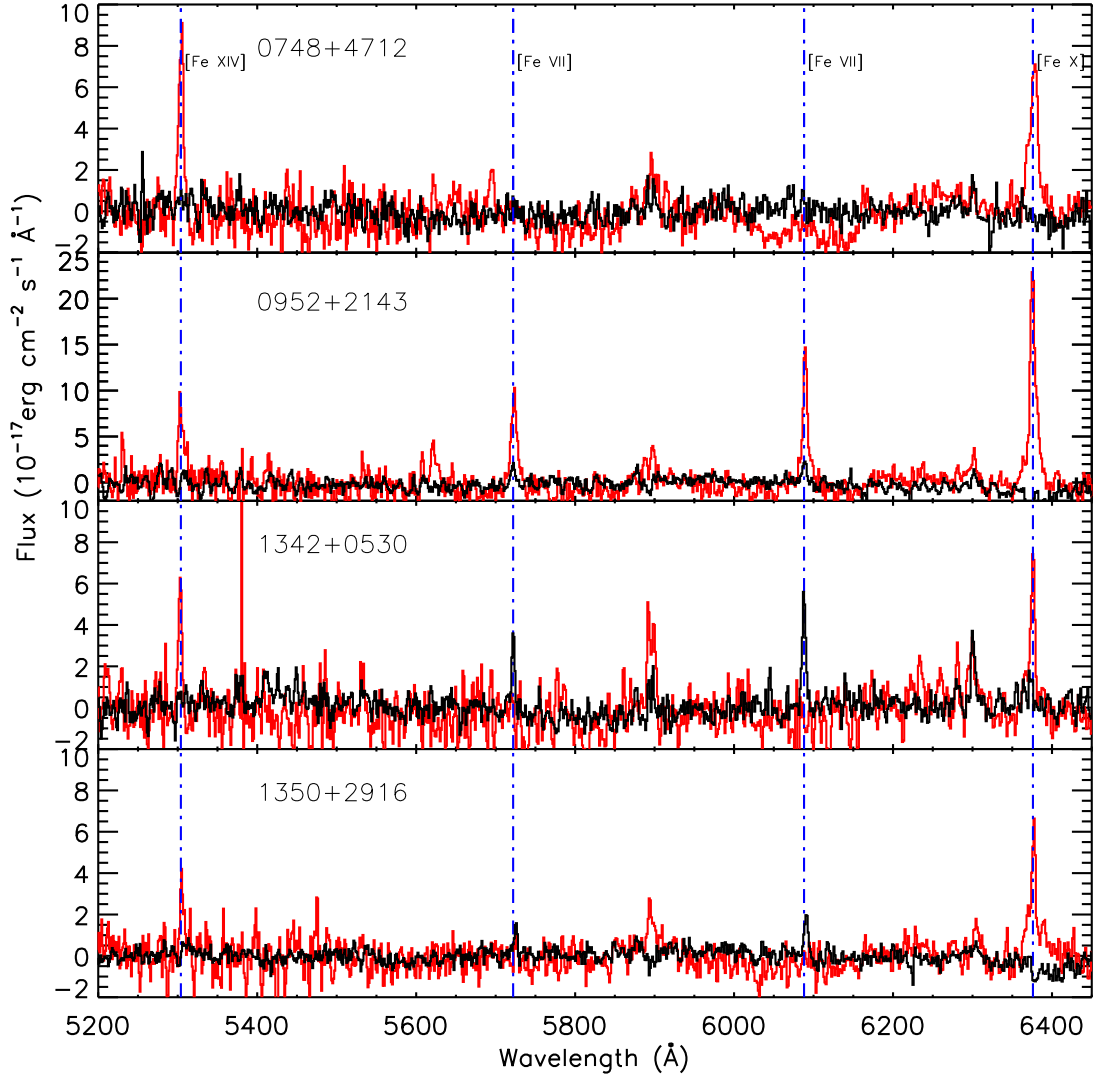}
\caption{Coronal line of four TDE targets. From top to bottom: J0748+4712, J0952+2143, J1342+0530, J1350+2916. (black: MMT spectra, red: SDSS spectra) }
\label{fig5}
\end{figure}  

\begin{figure}
\epsscale{0.8}
\plotone{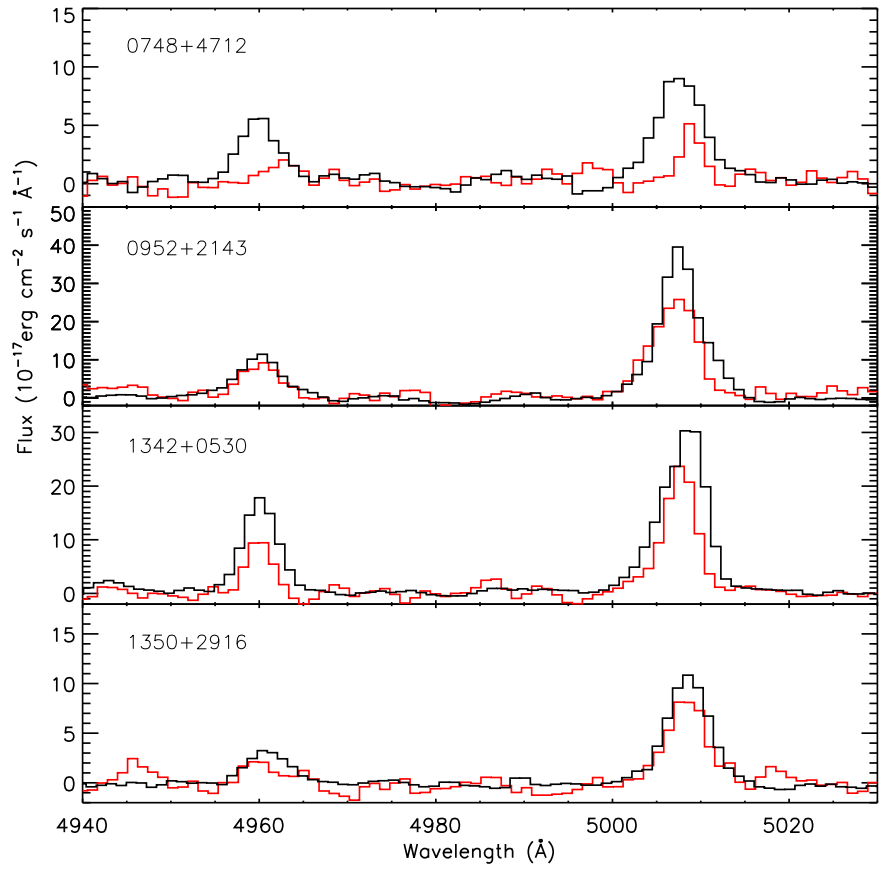}
\caption{[O {\sc III}] lines of four TDE targets. Same order and color as figure 5. }
\label{fig6}
\end{figure}  

\begin{figure}
\epsscale{0.5}
\plotone{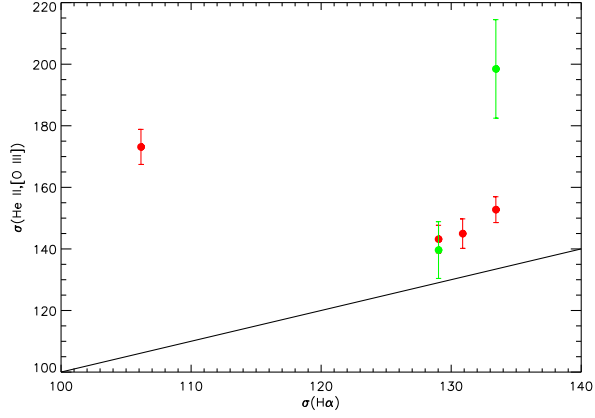}
\caption{Line width of \ha\ versus the width of [O {\sc III}] (red) and He II (green) line in four TDE targets in MMT spectra. Only lines with $>$ 3$\sigma$ detection are plotted. The straight line denotes one-to-one relation. }
\label{fig7}
\end{figure}  

\begin{figure}
\epsscale{1.0}
\plotone{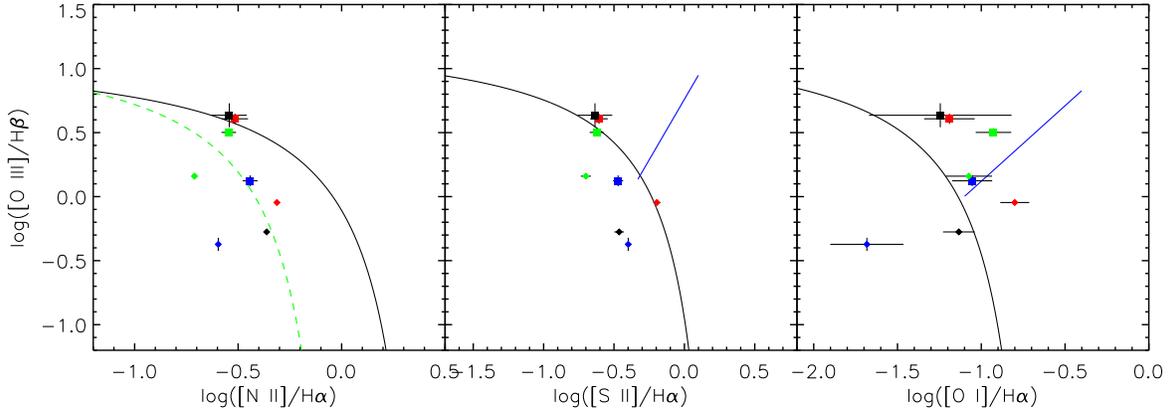}
\caption{Evolution of TDE targets in BPT diagrams (diamond: SDSS, square: MMT; blue: J0748+4712, red: J0952+2143, green: J1342+0530, black: J1350+2916). Those targets evolve from star-forming region towards extreme star-forming region and weak Seyfert region. The changes are mainly due to the increasing of [O {\sc III}] and smaller aperture of MMT spectra.}
\label{fig8}
\end{figure}

\begin{figure}
\epsscale{1.0}
\plotone{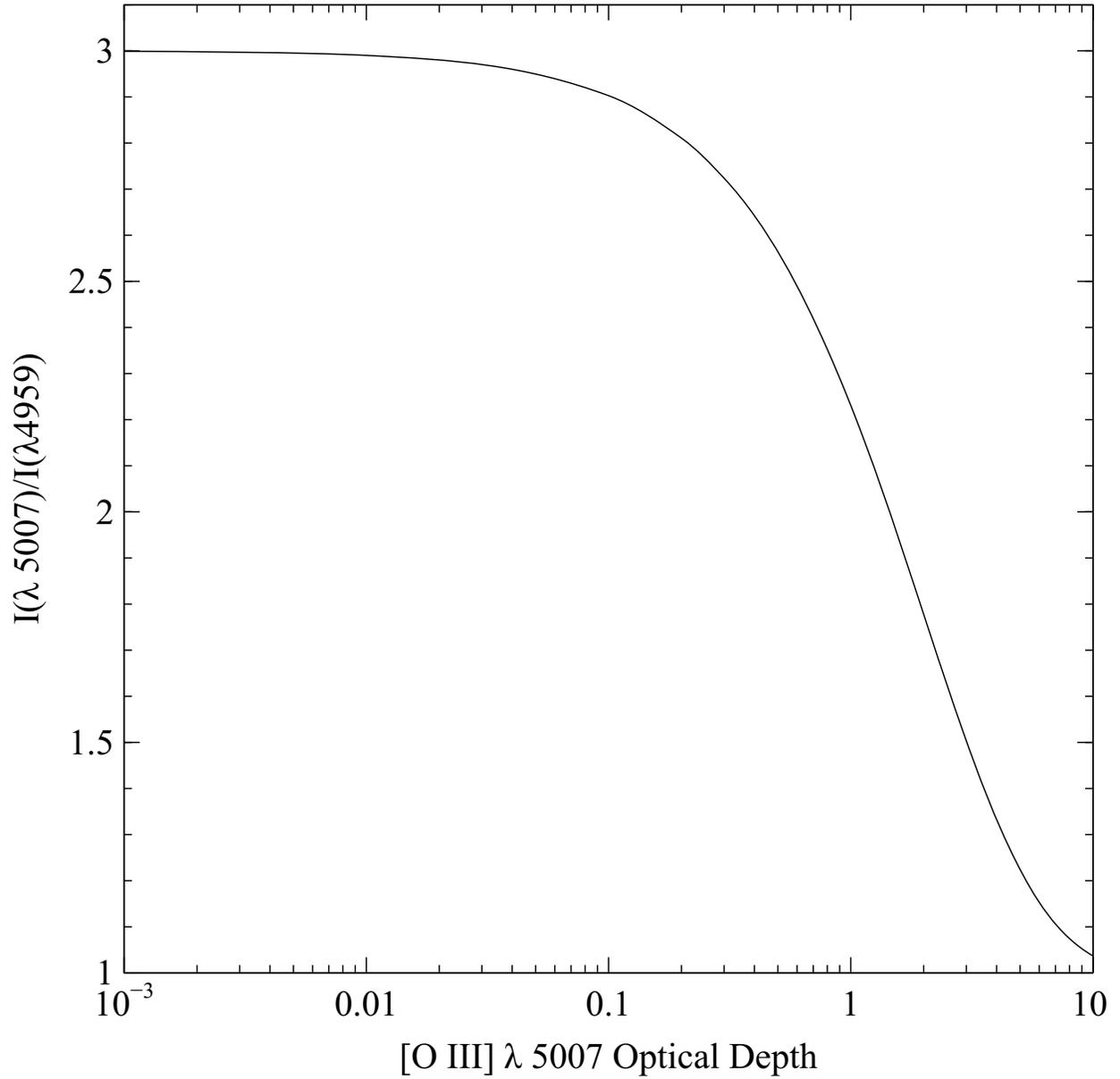}
\caption{The dependence of the [O {\sc III}] $\lambda 5007/\lambda 4959$ ratio on optical depth of $\lambda 5007$. A line ratio of $\approx 2$ corresponds to an optical depth $\tau(\lambda 5007) \approx 1.5$.}
\label{fig9}
\end{figure}

\begin{figure}
\epsscale{1.0}
\plotone{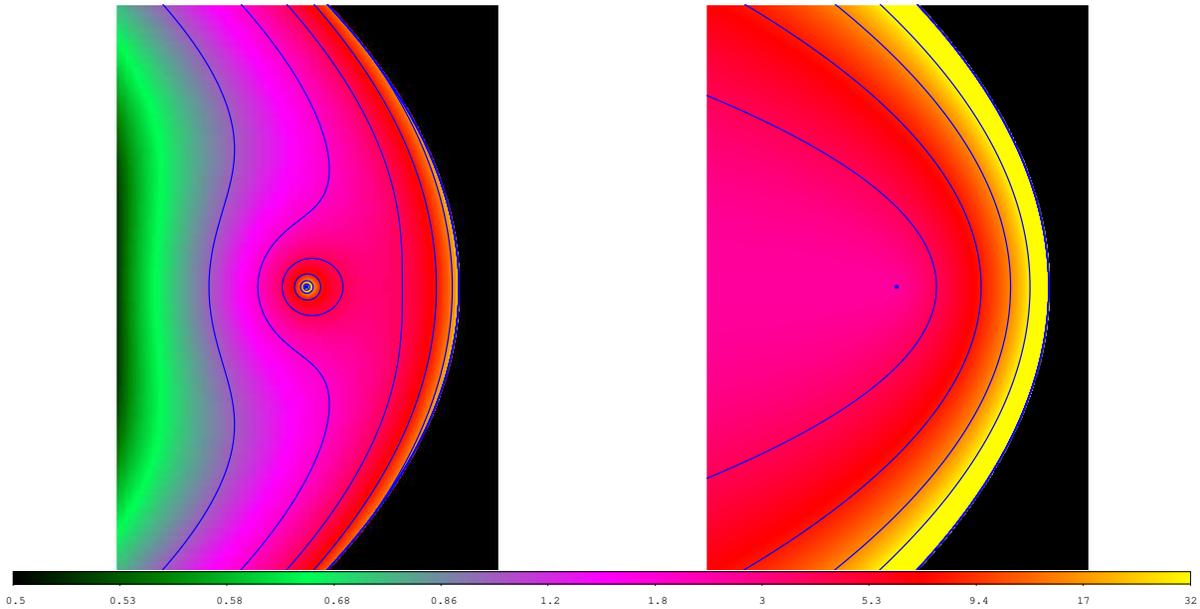}
\caption{A snapshot of the ionization parameter around a black hole that contributes to the observed emission lines at a specific time in an over-simplified model, in which the ionizing continuum
decreases with time as $\propto$ t$^{-5/3}$ when t $>$ t$_{0}$ (where t$_{0}$ is the time between the star was  tidal disrupted and most bound material returned to pericenter) and the density distribution decreases as $\propto$ r$^{-1}$(left panel) and $\propto$ r$^{-2}$(right panel). The observer is to the left of the figure at time 10 t$_{0}$. The overlaid blue lines are intensity contours with an interval of two times fold.}
\label{fig10}
\end{figure}
\end{document}